\documentstyle[12pt,epsf]{article}
\def\gt{\raisebox{0.2ex}{$>$}}

\begin{document}
\vspace*{0.2cm}
\rightline{\Large{DFPD 99/EP/4}}
\vspace*{0.3cm}
\rightline{January 25, 1999}
\begin{center}
\vspace*{.5 cm}
\Huge{\bf A real CKM matrix ?\\}
\vspace*{1.5cm}
\Large {P.Checchia\footnotemark[1],E. Piotto\footnotemark[2] and F. Simonetto\footnotemark[1]}\\
\footnotemark[1] {\it I.N.F.N sezione di Padova and Dipartimento di Fisica Universit\`{a} di Padova,
       \\ Padua, Italy\\ \rm}
\footnotemark[2]{\it I.N.F.N sezione di Milano and Dipartimento di Fisica Universit\`{a} di Milano,
       \\ Milan, Italy\\ \rm}
\vspace*{0.3cm}

\vspace*{0.3cm}
\end{center}
\begin{abstract}
The hypothesis of a  real Cabibbo-Kobayashi-Maskawa matrix has been considered and 
found to be disfavoured by present measurements even when neglecting results from
CP violation
in  neutral Kaon decay. This result contradicts statements 
reported in \cite{ref:mele}.
\end{abstract}
\newpage
\section{ Introduction}
The  Cabibbo-Kobayashi-Maskawa  \cite{ref:cabi,ref:koba} $3\times 3$ unitary matrix
\begin{equation}
V_{CKM} = \left(
\begin{array}{ccc}
V_{ud} & V_{us} & V_{ub} \\
V_{cd} & V_{cs} & V_{cb} \\
V_{td} & V_{ts} & V_{tb} \end{array}
\right)
\end{equation}

can be parameterized in terms of four parameters $\lambda,~A,~\rho$ and $\eta$ \cite{ref:wolf}:
\begin{equation}
V_{CKM} = \left(
\begin{array}{ccc}
1 - \lambda^{2}/2 & \lambda & A\lambda^{3}(\rho - i\eta) \\
-\lambda & 1 - \lambda^{2}/2  & A\lambda^{2} \\
A\lambda^{3}(1 - \rho - i\eta) & -A\lambda^{2} & 1 \end{array}
\right) + {\cal{O}}(\lambda^{4}).
\end{equation}

In this parameterization (or in its extention to $ {\cal{O}}(\lambda^{6})$ \cite{ref:paga}) 
the complex phase of the matrix is given by the parameter $\eta$ related to the CP
violation of the weak interactions. The $\lambda$ and $A$ parameters are known with
a good accuracy ($\sim 1 \%$ and $\sim 4 \%$, respectively) while  many contributions 
to extract $\rho$ and $\eta$ from the available measurements exist in the 
literature  \cite{ref:paga,ref:paro,ref:mele}.
To this purpose, the measurements of the CP violation parameter in neutral Kaon decay 
$\left| \epsilon_k \right|$, of the difference between the mass eigenstates in
the $B^0_d - \bar{B^0_d}$ system $\Delta m_d$
and of the ratio $\left| \frac{V_{ub}}{V_{cb}} \right|$ and the lower limit on the 
  difference between the mass eigenstates in
the $B^0_s- \bar{B^0_s}$ system $\Delta m_s$
can be used.
On the other hand, since the only direct experimental evidence for CP violation
is given by  the fact that $\left| \epsilon_k \right| \neq 0$ 
and the effect could be explained in terms of models proposed in
 alternative to the Standard Model (see, for instance,   \cite{ref:barb,ref:glas}
and references therein),
it is suggested to remove
the constraint coming from the neutral Kaon system and to investigate the   
results on parameter $\eta$
or to test the hypothesis of a real  $V_{CKM}$ matrix.
In this letter the latter hypothesis is tested with a different statistical approach 
with respect to \cite{ref:mele} coming to a different conclusion. In addition, the inclusion
of different data sets is discussed and the corresponding  results are presented.

\section {Measurements and constraints on $V_{CKM}$ parameters}

The $\lambda$ parameter is the sine of the Cabibbo angle 
\cite{ref:pdg}:
\begin{equation}
 \lambda =\left| V_{us} \right| =sin \theta_c = 0.2196 \pm 0.0023.
\end{equation}

The A parameter depends on the matrix element $\left| V_{cb} \right| = 0.0395 \pm 0.0017$
(obtained from semileptonic decays of B hadrons \cite{ref:pdg}) and on $\lambda$:

\begin{equation}
 A =\frac{\left| V_{cb} \right|}{\lambda^2} = 0.819 \pm 0.035.
\end{equation}

In order to constrain the parameters $\rho$ and $\eta$ without considering 
CP violation 
in the neutral Kaon system, three experimental input are used:

\subsection{ $B^0_d$ oscillations.}

 The mass difference $\Delta m_d$ between the mass eigenstates in the  
 $B^0_d - \bar{B^0_d}$ system has been measured with high precision
\cite{ref:pdg,ref:bosc}.
 In the Standard Model it can be related to the CKM parameters 
in the following way:

\begin{equation}
 \left[ (1-\rho)^2+\eta^2 \right] =
\frac{ \Delta m_d} 
 {\frac{G^2_F}{6\pi^2}m^2_t m_{B^0_d} 
 \left( f_{B_d} \sqrt{B_{B_d}}\right)^2 \eta_B F(z) A^2 \lambda^6 }
\end{equation}
where  $m_t$ is the
 the top pole mass scaled according to 
\cite{ref:buras} and $z=m^2_t/m^2_W$. The function $F(z)$ is given by:
\begin{equation}
    F(z) = \frac{1}{4} + \frac{9}{4(1-z)} - \frac{3}{2(1-z)^2} -
        \frac{3z^{2}\ln{z}} {2(1-z)^3}.
\end{equation}
The values of all the parameters are given in {\it table} \ref{tab:param}. 

In the  $\rho - \eta$ plane, the measurement of $\Delta m_d$ corresponds
to a circumference centered in (1,0). By constraining $\eta$  to zero,
an evaluation of $\rho$ can  be obtained. Unfortunately
the term $f_{Bd} \sqrt{B_{B_d}}$, given by lattice QCD calculations, 
is known with a  $20\%$ order  uncertainty  \cite{ref:fbd97,ref:fbd98} 
and  therefore  it gives the largest contribution to error on the $\rho$
determination.

\begin{table}[htb]
\centering
\begin{tabular}{|c|c|c|c|} \hline
parameter    & value used in \protect\cite{ref:mele}               & published data only& new input \\
\hline
 $G_F	$    &$1.16639(1)\times 10^{-5} \rm{GeV}^{-2}   $  & &	     \\
 $\lambda$   &$ 0.2196 \pm 0.0023  $                       & &            \\
 $A$         &$ 0.819 \pm 0.035    $                       & &            \\
 $m_t$       &$ 166.8 \pm 5.3      $ GeV                   & &            \\
 $m_W$       &$ 80.375 \pm 0.064    $GeV                   &$80.41 \pm 0.10  $GeV          &            \\
 $m_{B_d}$   &$ 5.2792 \pm 0.0018   $GeV                   & &            \\
 $m_{B_s}$   &$ 5.3692 \pm 0.0020   $GeV                   & &            \\
 $\eta_B$    &$ 0.55  \pm 0.01     $                       & &            \\
$\Delta m_d$ &$ 0.471 \pm 0.0016 ~\rm{ps}^{-1} $           
             &$0.464\pm 0.0018~\rm{ps}^{-1} $&            \\
$\Delta m_s$ &$ \gt 12.4~\rm{ps}^{-1}(95 \%~\rm{C.L.})   $ 
             &$ \gt 9.1~\rm{ps}^{-1}(95 \%~\rm{C.L.})   $ &            \\
$ \left|V_{ub} \right|/\left|V_{cb} \right|$&$0.093 \pm 0.016$ 
             &                              &$0.100 \pm 0.013$            \\
$f_{B_d} \sqrt{B_{B_d}}$ &$ 0.201 \pm 0.042$ GeV            
             &                           &$ 0.215^{+0.040}_{-0.030}$GeV \\
$ \xi$       &$ 1.14 \pm 0.08 $                            
             &                           &$ 1.14^{+0.07}_{-0.06} $      \\
\hline
\end{tabular}
\caption{ Physical parameters used in the formulae (5),(6),(7) and (8).
In the second column the values used in \protect\cite{ref:mele} and 
in the third column the values obtained from published data 
(or combinations quoted in \protect\cite{ref:pdg}) are given.
In the fourth column  new values for $f_{B_d} \sqrt{B_{B_d}}$ and $\xi$  
\protect\cite{ref:fbd98}
are given and
the preliminary DELPHI $V_{ub}$ measurement is included.
}
\label{tab:param}
\end{table}

\subsection{ $B^0_s$ oscillations.}
The mass difference $\Delta m_s$ between the mass eigenstates in the  
 $B^0_s - \bar{B^0_s}$ system  is expected 
to be much larger than  $\Delta m_d$ and in the Standard Model is related to
the CKM parameters:
\begin{equation}
 \left[ (1-\rho)^2+\eta^2 \right] =
 \frac{\Delta m_d} { \Delta m_s} \frac{1}{\lambda^2} \frac{ m_{B^0_s}}{ m_{B^0_d}} 
 \xi^2
\end{equation}

Since the ratio 
\begin{equation}
 \xi = \frac{  f_{B_s} \sqrt{B_{B_s}}} {f_{B_d} \sqrt{B_{B_d}}}
\end{equation}
is computed by lattice QCD with a better precision than the single terms,
a measurement of $\frac{ \Delta m_d}{ \Delta m_s} $ could provide a much 
stronger constraint on the $\rho - \eta$ plane. However, given the very high 
frequency in the  $B^0_s - \bar{B^0_s}$ system oscillation, 
only a lower limit on  $ \Delta m_s$ is available, 
as shown in {\it table} \ref{tab:param}, which  corresponds to a circular
bound  in the two parameter space or, if the assumption $\eta = 0$ is
made, to a lower limit for $\rho$.

\subsection {$ \left|V_{ub} \right|$ measurements from semileptonic b decay.}
 
Charmless semileptonic b decays have been used to measure $ \left|V_{ub} \right|$ 
or the ratio $ \left|V_{ub} \right| / \left|V_{cb} \right|$.
The CLEO collaboration determined that parameter both by measuring the rate 
of leptons produced in B semileptonic decays beyond the charm end-point \cite{ref:cleo} 
and
from direct reconstruction of charmless B semileptonic decay \cite{ref:cleo2}. 
The two results are consistent
but both methods are limited by theoretical uncertainties. In \cite{ref:pdg} 
they are not combined and a value $\left|V_{ub} \right|/\left|V_{cb} \right|=0.08\pm0.02 $
obtained from the former result is given.
At LEP, ALEPH \cite{ref:alvub},
 L3 \cite{ref:l3vub} and
more recently DELPHI  \cite{ref:devub} have measured the inclusive charmless semileptonic
transitions $b \rightarrow ul\nu$. The average value of the LEP measurements 
with the previous value is given in  {\it table} \ref{tab:param}. 
The ratio of the two CKM matrix elements is 
related to the $\rho$ and $\eta$ parameters by:

\begin{equation}
 \frac {\left|V_{ub} \right| }{ \left|V_{cb} \right|}= \lambda \sqrt{\rho^2 + \eta^2}
\end{equation}
and hence, in the  $\rho - \eta$ plane, 
the measurement of  $ \left|V_{ub} \right| / \left|V_{cb} \right|$ corresponds
to a circumference centered at the origin. If $\eta$ is assumed to be zero, it would
be proportional to  the $\rho$ absolute value.

\section {Data compatibility with a real CKM matrix hypothesis.}

The assumption of a real CKM matrix implies that all the constraints described
in the previous section are reduced to values of (or limits on) $\rho$.
The compatibility of the obtained values can then be used to estimate the goodness of the
assumption itself.
In  \cite{ref:mele} and in the explicit reference to it in  \cite{ref:glas},
it is written that the hypothesis of a real CKM matrix can fit the data.
However it is unclear 
which was the  statistical approach followed to come to that statement 
\footnote{In the cited paper, a result obtained by an underconstrained fit to 
$\rho$ and $\eta$ excludes the hypothesis $\eta=0$ at the $99 \%$ C.L. and another fit
with $\eta=0$ gives  $\chi^2=6.7$. Since in the fits there are
 bounded parameters clearly correlated with $\rho$ (i.e.  $f_{B_d} \sqrt{B_{B_d}}$) it is not
specified why the mentioned $\chi^2$ value corresponds to a reasonable Confidence 
Level for a real CKM matrix hypothesis.}
and a completely different
conclusion can be obtained with a standard method. 
In addition the claim of \cite{ref:mele}, although using a different input-data-set,
contradicts what is reported in \cite{ref:barb}. 
Assuming $\eta=0$ and using 
exactly the same input parameters as \cite{ref:mele} 
({\it table} \ref{tab:param} second column), the values

\begin{equation}
 \rho^{\Delta {m_d}}=0.01^{+0.18}_{-0.26}
\end{equation}
and
\begin{equation}
 \rho^{V_{ub}}=\pm (0.42 \pm 0.07)
\end{equation}

are obtained from eq. (5) and 
(9), respectively.

The limit on $\Delta m_s$  has been obtained by means of the
amplitude method \cite{ref:moser}
which allows to know the exclusion Confidence
Level  for any value of  $\Delta {m_s}$.
Therefore, by a convolution with the dominant uncertainty from the ratio $\xi$ in eq. (7)
it is possible to obtain: 
\begin{equation}
 \rho^{\Delta m_s}> -0.05 
\end{equation}
at the $95 \%$ Confidence Level.  

In a naive approach on which the errors on $\rho^{\Delta {m_d}}$ and $ \rho^{V_{ub}}$
are assumed to be uncorrelated
it is evident that the two values are fairly incompatible.
The negative $ \rho^{V_{ub}}$ solution is clearly excluded by the 
 $\rho^{\Delta m_s}$ limit and therefore it can be discarded.
In order to include all the correlations 
due to common terms contributing to the errors, namely $\left| V_{cb}\right|$ and $\lambda$,
a two dimensional error matrix $\bf M$ has been written and the 
Best Linear Unbiased Estimator
\cite{ref:blue} has been used:
\begin{equation}
 \rho^{BLUE}=\frac {\Sigma^2_{i=1}\Sigma^2_{j=1} \rho_i({\bf M}^{-1})_{ij} }
                   {\Sigma^2_{i=1}\Sigma^2_{j=1}       ({\bf M}^{-1})_{ij} }
\end{equation}  
with the variance
\begin{equation}
 \sigma^2_{\rho}=\frac {1 }
                   {\Sigma^2_{i=1}\Sigma^2_{j=1}       ({\bf M}^{-1})_{ij} }.
\end{equation}  
The error matrix $\bf M$ includes correlated and uncorrelated contributions:
\begin{equation}
M_{ij}=\delta_{ij}\sigma^{uncorr}_i\sigma^{uncorr}_j + \Sigma_{\alpha=1}^m \Delta_{\alpha i} \Delta_{\alpha j}
\end{equation}  

where the indexes $i$ and $j$ run over the two $\rho$ measurements and 
$\Delta_{\alpha i}$ is the change (with sign) on measurement $i$ when the common
systematic parameter $\alpha$ is moved by its error. 
The $\chi^2$ 
is  obtained by:
\begin{equation}
 \chi^2=
 \Sigma^2_{i=1}\Sigma^2_{j=1} \left( 
 \left[ \rho_i-\rho^{BLUE}\right] ({\bf M}^{-1})_{ij} \left[ \rho_j-\rho^{BLUE}\right]
 \right). 
\end{equation}

With the 
{\it table} \ref{tab:param} (second column)  parameters and taking the positive error 
in eq. (10),
\begin{equation}
   \rho^{BLUE}=(3.58 \pm 0.66) \times 10^{-1} ~ \rm{and} ~ \chi^2=4.3~ \rm{(1~ Degree~ of ~freedom)}
\end{equation}
are obtained. The corresponding  $\chi^2$ probability is $3.9 \%$ and this is 
clearly in contradiction with  \cite{ref:mele}. The correlation between the two measurements
is small given the dominance of the uncertainty on  $f_{B_d} \sqrt{B_{B_d}}$ 
in $\rho^{\Delta m_d}$. The effect
of the limit on  $\Delta {m_s}$ is negligible even though the information 
contained in the amplitude is taken into account as suggested in \cite{ref:paga}:
since the value of $\Delta {m_s}$ that one would obtain by inserting 
$\rho= \rho^{BLUE}$ and 
$\eta=0$ in eq. (7) is $\Delta {m_s}=31~ \rm{ps}^{-1} $, the present 
experimental sensitivity ($13.8~ \rm{ps}^{-1} $)
 does not give any sizeable information 
on that $\rho$ region.

Some of the parameters  used to determine $\rho$ and $\chi^2$ in eq. (16) have been 
obtained from preliminary results. It is then important to quantify the compatibility
of the real CKM hypothesis with published data. For such a test, the previous evaluation
is repeated with the parameters listed in  the third column of 
{\it table} \ref{tab:param} and

\begin{equation}
   \bar{\rho}^{BLUE}=(3.53 \pm 0.65) \times 10^{-1} ~\rm{and}
    ~\chi^2=4.0~{\rm(1~ Degree~ of~ freedom)}
\end{equation}
are obtained with a corresponding $\chi^2$ probability of $4.6 \%$.
Here, in order to take into account terms of the order up to 
$O(\lambda^5)$
 \cite{ref:buras2}, the substitution
  $\rho \rightarrow \bar{\rho}=\rho (1-\lambda^2/2) $
has been done in eq. (5) 
with $\eta=0$.
This result, given the little change on parameters, is similar to the previous one
and therefore the hypothesis of a real CKM matrix is disfavoured also by published data.

If a more recent value of $f_{B_d} \sqrt{B_{B_d}}$ is taken and the DELPHI $V_{ub}$ preliminary
result is included (see {\it table} \ref{tab:param}), the incompatibility of the two measurements
is still present:

\begin{equation}
   \bar{\rho}^{BLUE}=(3.91 \pm 0.53) \times 10^{-1} 
   ~\rm{and}~\chi^2=4.6~ {\rm(1~ Degree~ of~ freedom)}
\end{equation}
corresponding to a  $\chi^2$ probability of $3.1 \%$.    
 
Since the dominant error 
in $\rho^{\Delta m_d}$ is due to the lattice QCD computation, the hypothesis of
a flat error with the same R.M.S. on $f_{B_d} \sqrt{B_{B_d}}$ has been studied 
with a simple simulation. Several experimental results for 
$\rho^{\Delta m_d}$ and $\rho^{V_{ub}}$
have been generated 
with central values equal to $\rho^{BLUE}$. For $\rho^{\Delta m_d}$  this is achieved
by shifting the value of $f_{B_d} \sqrt{B_{B_d}}$ and allowing it to vary within
a flat distribution  with the R.M.S. corresponding to the quoted error. All the other 
parameters of eq. (5)
are allowed to vary with a gaussian distribution
corresponding to their error. In each experiment the combined value and the
$\chi^2$ are computed according to eq. (13) and (16), respectively.
 Looking at the $\chi^2$ 
 distribution, it is possible
to determine the fraction of simulated experiments with a 
$\chi^2$ higher than the
value found in the evaluation with real data.
This procedure has been repeated for the three data sets of {\it table} \ref{tab:param}
and for none of them the $\chi^2$ has been found to be higher than the
experimental values of eq. (17), (18) and (19) in more than $5\%$ of the cases.

\section{Conclusions}

The hypothesis of a real CKM matrix is tested on the basis of the present published
and preliminary data and lattice QCD calculations. With all the input data used,
included those suggested in \cite{ref:mele}, that hypothesis is excluded at more than
$95 \%$ Confidence Level. This result contradicts statements in \cite{ref:mele} and
agrees with the result indicated in \cite{ref:barb} with an older input data set.

\subsection*{Acknowledgements}
We whish to thank M. Loreti, G. Martinelli, M. Mazzucato, F.Parodi and A.Stocchi for 
comments and useful discussions.





\begin{thebibliography}{99}

\bibitem{ref:mele} S. Mele, CERN-EP/98-133 submitted to {\em Phys. Lett. B}.
\bibitem{ref:cabi} N. Cabibbo, {\em Phys. Rev. Lett.} {\bf 10} (1963) 531.
\bibitem{ref:koba} M. Kobayashi and T. Maskawa, {\em Prog. Theor. Phys.} {\bf 49} (1973) 652.
\bibitem{ref:wolf} L. Wolfenstein, {\em Phys. Rev. Lett.} {\bf 51} (1983) 1945.
\bibitem{ref:paga} P. Paganini et al., {\em Phys. Scripta} {\bf 58} (1998) 556.
\bibitem{ref:paro} F. Parodi et al. hep-ph/9802289.
\bibitem{ref:barb} R. Barbieri et al., {\em  Phys.  Lett.} {\bf B 425} (1998) 119.
\bibitem{ref:glas} H. Georgi and S. Glashow, HUTP-98/A048.
\bibitem{ref:pdg} Particle Data Group, {\em Eur. Phys. J.} {\bf C 3} (1998) 1.
\bibitem{ref:bosc} The LEP B Oscillation working group, LEPBOSC 98/3
    contribution to the Vancouver 1998 conference.
\bibitem{ref:fbd97} J.M. Flynn and C.T. Sachraida, hep-lat/9710057,
 Proceedings of: {\em 4th International Workshop on Progress in Heavy Quark Physics} Rostock, Germany.
\bibitem{ref:fbd98} T. Draper, hep-lat/9810065,
{\em $14^{th}$ International Symposium on Lattice Field Theory: Lattice '98} Boulder, CO, UK .
\bibitem{ref:buras} A. J. Buras et al., {\em Nucl. Phys.} {\bf B 347} (1990) 491.
\bibitem{ref:cleo} CLEO collab., J.Bartelt et al., {\em  Phys. Rev. Lett.} {\bf 71} (1993) 4111.
\bibitem{ref:cleo2} CLEO collab, J.P. Alexander et al., {\em  Phys. Rev. Lett.} {\bf 77} (1996) 5000.
\bibitem{ref:alvub} ALEPH collab., R. Barate et al., CERN-EP/98-067 accepted by 
                      {\em Euro. Phys. J.}{\bf C}.
\bibitem{ref:l3vub} L3 collab., M. Acciarri et al., {\em  Phys.  Lett.} {\bf B 436} (1998) 174.
\bibitem{ref:devub} DELPHI collab., M. Battaglia et al. DELPHI 98-97 CONF 165.
\bibitem{ref:moser} H.G. Moser and A. Roussarie, {\em Nucl. Instr. and Methods}
                  {\bf A 384} (1997) 491.
\bibitem{ref:blue} See for example L.Lyons at al. {\em Nucl. Instr. and Methods} 
                   {\bf A 270} (1988) 110 and references therein; 
details of the method are also described in COMBOS manual
available at http://www.cern.ch/LEPBOSC/combos. 
\bibitem{ref:buras2} A. J. Buras et al., {\em Phys. Rev.} {\bf D 50} (1994) 3433.
\end{thebibliography}
\end{document}